\documentclass[12pt]{article}
\usepackage{amsmath,amssymb,bm,graphicx}
\usepackage{color} 

\newcommand{\pslash}{\not \! p}

\newcommand{\delslash}{\not \! \partial}

\usepackage{hyperref}
\usepackage{cite}

\begin{document}

%\vskip 0.5 truecm

\begin{center}
{\Large{\bf Baryon number violation and novel canonical anti-commutation relations }}
\end{center}\vskip .5 truecm
\begin{center}
{\bf { Kazuo Fujikawa$^{1,2}$ and Anca Tureanu$^1$ }}
\end{center}

\begin{center}
\vspace*{0.4cm} 
{\it {$^1$Department of Physics, University of Helsinki, P.O.Box 64, 
\\FIN-00014 Helsinki,
Finland\\
$^2$Quantum Hadron Physics Laboratory, RIKEN Nishina Center,\\
Wako 351-0198, Japan
}}
\end{center}

\begin{abstract} 
The possible neutron-antineutron oscillation is described by an effective quadratic Lagrangian analogous to the BCS theory. It is shown that the conventional equal-time anti-commutation relations of the neutron variable $n(t,\vec{x})$ are modified by the baryon number violating terms. This is established by the Bjorken--Johnson--Low prescription and also by the canonical quantization combined with equations of motion. This novel canonical behavior can give rise to an important physical effect, which is illustrated by  analyzing the Lagrangian that violates the baryon number but gives rise to the degenerate effective Majorana fermions and thus no neutron-antineutron oscillation.  Technically, this model is neatly treated using a relativistic analogue of the Bogoliubov transformation.   
\end{abstract}

%\maketitle
%\large
\section{Introduction}

The possible  neutron oscillation is analyzed by the quadratic effective Hermitian Lagrangian with general $\Delta B=2$ terms added \cite{kuzmin, mohapatra, glashow, chang1, marshak, kuo, kazarnovsky, mohapatra2,chang,ft,nelson,gardner},
\begin{eqnarray}\label{1}
{\cal L}_{0}&=&\overline{n}(x)i\gamma^{\mu}\partial_{\mu}n(x) - m\overline{n}(x)n(x)\nonumber\\
&-&\frac{1}{2}\epsilon_{1}[e^{i\alpha}n^{T}(x)Cn(x) + e^{-i\alpha}\overline{n}(x)C\overline{n}^{T}(x)]\nonumber\\
&-&\frac{1}{2}\epsilon_{5}[n^{T}(x)C\gamma_{5}n(x) - \overline{n}(x)C\gamma_{5}\overline{n}^{T}(x)],
\end{eqnarray}
where $m$, $\epsilon_{1}$, $\epsilon_{5}$ and $\alpha$ are real parameters. The most general quadratic hermitian Lagrangian is written in the form \eqref{1} using the phase freedom of $n(x)\rightarrow n(x)=e^{i\beta}n^{\prime}(x)$; under this change of naming the field, the physical quantities
in \eqref{1} such as mass eigenvalues are obviously invariant. But C (and thus CP) transformation rules  of the solution of the Lagrangian \eqref{1} are modified. In the present paper, we adopt the above phase convention which is different from the one used in \cite{FT1}. 

The first $\Delta B=2$ term with real $\epsilon_{1}$ breaks the $\gamma^{0}$-parity which is defined by 
\begin{eqnarray}\label{2}
n(t,\vec{x}) \rightarrow \gamma^{0}n(t,-\vec{x}), \ \ \ n^{c}(t,\vec{x}) \rightarrow -\gamma^{0}n^{c}(t,-\vec{x})
\end{eqnarray}
with $n^{c}(t,\vec{x})\equiv C\overline{n(t,\vec{x})}^{T}$,
while the second term with real $\epsilon_{5}$ preserves $\gamma^{0}$-parity. In contrast, the first  term with real $\epsilon_{1}$ preserves $i\gamma^{0}$-parity which is defined by 
\begin{eqnarray}\label{3}
n(t,\vec{x}) \rightarrow i\gamma^{0}n(t,-\vec{x}), \ \ \ n^{c}(t,\vec{x}) \rightarrow i\gamma^{0}n^{c}(t,-\vec{x}),
\end{eqnarray}
while the second term with real $\epsilon_{5}$ breaks $i\gamma^{0}$-parity. The $i\gamma^{0}$-parity is natural to analyze the Majorana fermion since it preserves the reality of the field in the Majorana representation. In the discussion of discrete symmetries of the general effective Lagrangian \eqref{1}, one is bound to adopt the $i\gamma^{0}$-parity, and the CP defined in terms of $i\gamma^{0}$-parity is broken only when $\alpha\neq 0$ in \eqref{1}. Our notational conventions follow \cite{bjorken}, in particular, $C=i\gamma^{2}\gamma^{0}$. 

The model \eqref{1}  has been studied by various authors in the past \cite{kuzmin, mohapatra, glashow, chang1, marshak, kuo, kazarnovsky, mohapatra2,chang,ft,nelson,gardner}. We have given an exact solution of \eqref{1} with $\alpha\neq 0$ and showed that the neutron oscillation cannot detect the effect of CP violation, although the absolute rate of the oscillation is influenced by $\alpha\neq 0$~\cite{FT1}. We have also shown that the choice $\epsilon_{1}=0$ gives rise to the degenerate effective Majorana masses and thus no oscillation. Nevertheless, physically the effect of $\gamma_0$-parity preserving $\Delta B=2$ terms is not negligible~\cite{FT1}, and it may appear in the instability of nuclei. This effect is related to the interesting novel anti-commutation relations of neutron variables such as $\{n(t,\vec{x}),n(t,\vec{y})\}=0$ but $\{\dot{n}(t,\vec{x}),n(t,\vec{y})\}\neq 0$, which is analyzed in detail in the present paper. This effect is specific to the baryon number violating theory. 

For example, in a model analogous to the Nambu--Jona-Lasinio model~\cite{nambu}  such as
\begin{eqnarray}\label{NJL}
{\cal L}&=&\overline{n}(x)i\gamma^{\mu}\partial_{\mu}n(x) - m\overline{n}(x)n(x)\nonumber\\
&-&\lambda\overline{n}(x)(1+\gamma_{5})n(x)\overline{n}(x)(1-\gamma_{5})n(x),
\end{eqnarray}
where the baryon number is strictly conserved and thus
\begin{eqnarray}\label{nambu}
\langle T^{\star}n(x)n(y)\rangle=0,
\end{eqnarray}
the above mentioned novel behavior of the canonical anti-commutation relations does not appear.

\section{Degenerate Majorana masses}
We have shown that the effective Lagrangian \eqref{1} with $\epsilon_{1}=0$, i.e.
\begin{eqnarray}\label{4}
{\cal L}&=&\overline{n}(x)i\gamma^{\mu}\partial_{\mu}n(x) - m\overline{n}(x)n(x)\nonumber\\
&-&\frac{1}{2}\epsilon_{5}[n^{T}(x)C\gamma_{5}n(x) - \overline{n}(x)C\gamma_{5}\overline{n}^{T}(x)],
\end{eqnarray}
which is invariant under the "$\gamma^{0}$-parity",
gives rise to the degenerate Majorana fermions~\cite{FT1}, even in a more general context, as is discussed  later. 
The degeneracy of Majorana masses implies the absence of the conventional neutron oscillation despite the presence of the $\epsilon_{5}$-term with $\Delta B=2$. 

We use the Lagrangian in \eqref{4} to analyze the novel anti-commutation relations. To solve \eqref{4}, we apply an analogue of Bogoliubov transformation, $(n, n^{c})\rightarrow (N, N^{c})$, defined as~\cite{FT1}
\begin{eqnarray}\label{10}
\left(\begin{array}{c}
            N(x)\\
            N^{c}(x)
            \end{array}\right)
&=& \left(\begin{array}{c}
            \cos\phi\, n(x)-\gamma_{5}\sin\phi\, n^{c}(x)\\
            \cos\phi\, n^{c}(x)+\gamma_{5}\sin\phi\, n(x)
            \end{array}\right),
\end{eqnarray}
with
\begin{eqnarray}\label{mixing}
\sin 2\phi =\epsilon_{5}/\sqrt{m^{2}+(\epsilon_{5})^{2}}.
\end{eqnarray}
One can confirm the classical consistency condition $N^{c}=C\overline{N}^{T}(x)$ using the expressions of the right-hand side of \eqref{10}.
One can also confirm
\begin{eqnarray}
{\cal L}&=&\frac{1}{2}\{\bar{N}i\delslash N + \bar{N^{c}}i\delslash N^{c}\}\nonumber\\
&=&\frac{1}{2}\{\bar{n}i\delslash n + \bar{n^{c}}i\delslash n^{c}\}.
\end{eqnarray}
We can then show that the anticommutators are preserved, i.e.,
\begin{eqnarray}\label{anti-comm}
&& \{N(t,\vec{x}), N^{c}(t,\vec{y})\}=\{n(t,\vec{x}), n^{c}(t,\vec{y})\},\nonumber\\  
&&\{N_{\alpha}(t,\vec{x}), N_{\beta}(t,\vec{y})\}=\{N^{c}_{\alpha}(t,\vec{x}), N^{c}_{\beta}(t,\vec{y})\}=0,
\end{eqnarray}  
and thus the condition of a canonical trasformation required for the Bogoliubov transformation is satisfied. This condition of the canonical transformation is valid irrespective of the mass values of $n$ and $N$.  A transformation analogous to \eqref{10} has been  successfully used in the analysis of neutrino masses in the seesaw mechanism \cite{FT, FT2}.

After the Bogoliubov transformation, \eqref{4} becomes
\begin{eqnarray}\label{13}
{\cal L}&=&\frac{1}{2}\left[\overline{N}(x)\left(i\delslash - M\right)
 N(x)+\overline{N^{c}}(x)\left(i\delslash - M\right)N^{c}(x)\right]\nonumber\\
 &=&\frac{1}{2}\left[\overline{\psi_{+}}(x)\left(i\delslash - M\right)
 \psi_{+}(x)+\overline{\psi_{-}}(x)\left(i\delslash - M\right)\psi_{-}(x)\right],
\end{eqnarray}
where the Majorana fermions are defined by
\begin{eqnarray}\label{majorana-fermion}
\psi_{\pm}(x)=\frac{1}{\sqrt{2}}[N(x)\pm N^{c}(x)]
\end{eqnarray}
which satisfy
\begin{eqnarray}
\psi_{+}^{c}(x)=\psi_{+}(x), \ \ \ \psi^{c}_{-}(x)=-\psi_{-}(x).
\end{eqnarray}
The mass parameter is defined by
\begin{eqnarray}\label{14}
M\equiv \sqrt{m^{2}+(\epsilon_{5})^{2}}.
\end{eqnarray}
This implies that the Bogoliubov transformation maps the original theory to
a theory of quasiparticles described by the field $N(x)$, characterized by a new mass  $M$ ($\epsilon_{5}$ corresponds to the energy gap). The Bogoliubov  transformation maps a linear combination of a Dirac fermion and its charge conjugate to another Dirac fermion and its charge conjugate, and thus the Fock vacuum is mapped to a new  vacuum defined by ${\cal L}$ at $t=0$ (see, for example,\cite{FT2}).  It is important that the Bogoliubov transformation \eqref{10} preserves the CP symmetry, although it does not preserve the transformation properties under $i\gamma^{0}$-parity and C separately.

The solution of the starting Lagrangian \eqref{4} is written as,
\begin{eqnarray}\label{12}
\left(\begin{array}{c}
            n(x)\\
            n^{c}(x)
            \end{array}\right)
&=& \left(\begin{array}{c}
            \cos\phi N(x)+\gamma_{5}\sin\phi N^{c}(x)\\
            \cos\phi N^{c}(x)-\gamma_{5}\sin\phi N(x)
            \end{array}\right),
\end{eqnarray}
with $\sin2\phi$ defined in \eqref{mixing}.
The solution can also be expressed in terms of Majorana fermions defined in \eqref{majorana-fermion} using
\begin{eqnarray}
N(x)&=&[\psi_{+}(x)+ \psi_{-}(x)]/\sqrt{2}\nonumber\\
N^{c}(x)&=&[\psi_{+}(x)- \psi_{-}(x)]/\sqrt{2}.
\end{eqnarray}

When one generates the neutron experimentally, one obtains the field expressed as
\begin{eqnarray}\label{solution}
n(x)&=&\cos\phi N(x)+\gamma_{5}\sin\phi N^{c}(x)\nonumber\\
&=&\frac{1}{\sqrt{2}}\{\cos\phi [\psi_{+}(x)+ \psi_{-}(x)]+\gamma_{5}\sin\phi [\psi_{+}(x)- \psi_{-}(x)]\},\nonumber\\
n^{c}(x)&=&\cos\phi N^{c}(x)-\gamma_{5}\sin\phi N(x)\nonumber\\
&=&\frac{1}{\sqrt{2}}\{\cos\phi [\psi_{+}(x)- \psi_{-}(x)]-\gamma_{5}\sin\phi [\psi_{+}(x)+ \psi_{-}(x)]\},
\end{eqnarray}
but no oscillation in the conventional sense 
\begin{eqnarray}
n(x) \rightarrow n^{c}(x) \rightarrow n(x) \rightarrow....,
\end{eqnarray}
takes place due to the mass degeneracy of the Majorana fermions $\psi_{\pm}(x)$. Note that the neutron-antineutron oscillation $n(x) \rightarrow n^{c}(x)$ occurs due to the mass differences of the two Majorana particles appearing in the expressions of $n(x)$ and $n^{c}(x)$. It may thus appear that  no physical effects
of the baryon number violation such as the decay originating from $n(x)$ into two distinct final states appear. 

However, $n(x)$ and $n^{c}(x)$ are not orthogonal, in the sense 
\begin{eqnarray}\label{15}
\langle T^{\star} n^{c}(x)\bar{n}(y)\rangle
= \int\frac{d^{4}p}{(2\pi)^{4}}\frac{(-i)\gamma_{5}M\sin2\phi}{p^{2}-M^{2}+i\epsilon}e^{-ip(x-y)},
\end{eqnarray}
which is obtained from \eqref{solution} and the relations valid in \eqref{13},
\begin{eqnarray}\label{propagator}
\langle T^{\star} N(x)\overline{N}(y)\rangle&=&\langle T^{\star} N^{c}(x)\overline{N^{c}}(y)\rangle
= \int\frac{d^{4}p}{(2\pi)^{4}}\frac{i}{\pslash-M+i\epsilon}e^{-ip(x-y)}.
\end{eqnarray}
The relation \eqref{15}  shows that the propagation 
\begin{eqnarray}
n(x)\rightarrow n^{c}(y)
\end{eqnarray}
is possible, and thus  $n(x)$ can decay through 
\begin{eqnarray}
n\rightarrow p+e+\bar{\nu}_{e}, \ \ \ {\rm or}\ \ \ n^{c}\rightarrow \bar{p}+e^{+}+\nu_{e},
\end{eqnarray}
and the neutron annihilates when it collides with the ordinary matter containing the neutron, or by dinucleon decay within nuclei. Yet the oscillation is absent, which implies the absence of "bunching effect". The bunching effect here means that one would observe predominantly $n^{c}(x)$
starting with $n(x)$, when observed at a proper moment after the creation of $n(x)$, if the neutron-antineutron oscillation takes place. Various experiments have been searching for neutron-antineutron conversion both with free neutron beams, but also within nuclei. The present experimental status can be found in \cite{exp}.

Finally, we comment on the basic mechanism which generates the degenerate 
Majorana fermions.  
The "parity-doublet theorem" which was analyzed in \cite{FT1} states that the effective quadratic Lagrangian, if invariant under  the "$\gamma^{0}$-parity", gives rise to solutions which belong to the well-defined representations of the "$\gamma^{0}$-parity". If the solution is a superposition of $n(x)$ or $n^{c}(x)$ and thus cannot be the eigenstates of "$\gamma^{0}$-parity", the two possible solutions $\psi_{+}(t,\vec{x})$ and $\psi_{-}(t,\vec{x})$ form a {\em doublet representation},
\begin{eqnarray}
\psi_{+}(t,\vec{x})\rightarrow \gamma^{0}\psi_{-}(t,-\vec{x}), \ \ \ \psi_{-}(t,\vec{x})\rightarrow \gamma^{0}\psi_{+}(t,-\vec{x}).
\end{eqnarray}
Note that this representation of parity satisfies $P^{2}=1$. This parity doublet theorem combined with the equation of motion such as 
\begin{eqnarray}
{[}i\delslash -M]\psi_{+}(x)=0
\end{eqnarray}
implies that two Majorana-type fermions are degenerate in mass.  

%Note that P and CP of \eqref{4} are defined in terms of "$i\gamma^{0}$-parity" which is natural to define Majorana fermions.

\section{Novel canonical anti-commutation relations}
We want to show that the baryon number violating theory in general has an interesting novel property in the canonical anti-commutation relations, which, to our knowledge, have not been discussed before.

\subsection{Bjorken--Johnson--Low prescription}
We analyze the specific example in \eqref{4} by first using the  Bjorken--Johnson--Low (BJL) prescription, which is convenient to convert the results of path integrals (or propagator theory in general) to those in canonical quantization. We shall present a conventional canonical analysis later.
 An interesting consequence of the relativistic Bogoliubov transformation is that we have \eqref{15},
 \begin{eqnarray}\label{Tstar-product}
\int d^{4}x e^{ip(x-y)}\langle T^{\star} n^{c}(x)\bar{n}(y)\rangle
= \frac{(-i)\gamma_{5}M\sin2\phi}{p^{2}-M^{2}+i\epsilon}.
\end{eqnarray}

The basic machinery to analyze this correlation is the BJL prescription. This prescription states that one can replace the covariant $T^{\star}$ product, which does not specify the equal-time limit precisely, by the conventional $T$ product, which specifies the equal-time limit of the correlation precisely, if the correlation specified by $T^{\star}$ vanishes for $p^{0}\rightarrow \infty$.  In concrete terms, for two arbitrary operators $A(x)$ and $B(x)$, if
\begin{equation}\label{BJL_condition}
\lim_{p_0\to\infty}\int d^{4}x e^{ip(x-y)}\langle T^{\star} A(x)B(y)\rangle=0
\end{equation}
then
\begin{equation}
\int d^{4}x e^{ip(x-y)}\langle T^{\star} A(x)B(y)\rangle=\int d^{4}x e^{ip(x-y)}\langle T A(x)B(y)\rangle.
\end{equation}
If \eqref{BJL_condition} is not fulfilled, one defines the $T$-product by subtraction:
\begin{eqnarray}\label{T-prod_def}
\int d^{4}x e^{ip(x-y)}\langle T A(x)B(y)\rangle&=&\int d^{4}x e^{ip(x-y)}\langle T^{\star} A(x)B(y)\rangle\cr
&-&\lim_{p_0\to\infty}\int d^{4}x e^{ip(x-y)}\langle T^{\star} A(x)B(y)\rangle,
\end{eqnarray}
thus ensuring that the limit $p_0\to\infty$ of any $T$-product of operators vanishes.

This criterion is regarded as an analogue~\cite{fujikawa-suzuki} of Riemann--Lebesgue lemma in Fourier transform: if a function $f(t)$ is smooth and well-defined around $t=0$, the large frequency limit $\omega\rightarrow \infty$ of $\int_{-\infty}^{\infty} dt e^{i\omega t}f(t)$ vanishes.

One can confirm that \eqref{Tstar-product} satisfies this condition. We thus obtain 
\begin{eqnarray}\label{T-product}
\int d^{4}x e^{ip(x-y)}\langle T n^{c}(x)\bar{n}(y)\rangle
= \frac{(-i)\gamma_{5}M\sin2\phi}{p^{2}-M^{2}+i\epsilon}.
\end{eqnarray}
We next multiply both sides by $p^{0}$, and we obtain
\begin{eqnarray}\label{30x}
&&p^{0}\int d^{4}x e^{ip(x-y)}\langle T n^{c}(x)\bar{n}(y)\rangle\nonumber\\
&=&\int d^{4}x e^{ip(x-y)}i\partial_{x^{0}}\langle T n^{c}(x)\bar{n}(y)\rangle\nonumber\\
&=&\int d^{4}x e^{ip(x-y)}[i\delta(x^{0}-y^{0})\{n^{c}(t,\vec{x}),\bar{n}(t,\vec{y})\}+
\langle T i\partial_{x^{0}}n^{c}(x)\bar{n}(y)\rangle]\nonumber\\
&=& \frac{p^{0}(-i)\gamma_{5}M\sin2\phi}{p^{2}-M^{2}+i\epsilon}.
\end{eqnarray}
where we used 
\begin{eqnarray}
\partial_{x^{0}}\langle T n^{c}(x)\bar{n}(y)\rangle=\delta(x^{0}-y^{0})\{n^{c}(t,\vec{x}),\bar{n}(t,\vec{y})\}+
\langle T \partial_{x^{0}}n^{c}(x)\bar{n}(y)\rangle.
\end{eqnarray}
Next, we take the limit $p^{0}\rightarrow \infty$ in \eqref{30x}, obtaining:
\begin{eqnarray}\label{30x}
\lim_{p_0\to\infty}\int d^{4}x e^{ip(x-y)}[i\delta(x^{0}-y^{0})\{n^{c}(t,\vec{x}),\bar{n}(t,\vec{y})\}+
\langle T i\partial_{x^{0}}n^{c}(x)\bar{n}(y)\rangle]=0.
\end{eqnarray}
The second term will vanish, by definition of the $T$-product (see eq. \eqref{T-prod_def}). The first term, owing to the presence of $\delta(x^{0}-y^{0})$, is independent of $p_0$. Thus we infer
\begin{eqnarray}
i\delta(x^{0}-y^{0})\{n^{c}(t,\vec{x}),\bar{n}(t,\vec{y})\}=0.
\end{eqnarray}
Returning with this result into \eqref{30x}, we obtain also 
\begin{eqnarray}\label{first-step}
\int d^{4}x e^{ip(x-y)}
\langle T i\partial_{x^{0}}n^{c}(x)\bar{n}(y)\rangle
&=& \frac{p^{0}(-i)\gamma_{5}M\sin2\phi}{p^{2}-M^{2}+i\epsilon}.
\end{eqnarray}
We next multiply by $p^{0}$ both sides of \eqref{first-step}:
\begin{eqnarray}
&&p^{0}\int d^{4}x e^{ip(x-y)}
\langle T i\partial_{x^{0}}n^{c}(x)\bar{n}(y)\rangle\nonumber\\
&=&\int d^{4}x e^{ip(x-y)}i\partial_{x^{0}}
\langle T i\partial_{x^{0}}n^{c}(x)\bar{n}(y)\rangle\nonumber\\
&=&\int d^{4}x e^{ip(x-y)}\Big[i\delta(x^{0}-y^{0})\{i\partial_{x^{0}}n^{c}(t,\vec{x}),\bar{n}(t,\vec{y})\}+
\langle T (i\partial_{x^{0}})^{2}n^{c}(x)\bar{n}(y)\rangle\Big]\nonumber\\
&=& \frac{(p^{0})^{2}(-i)\gamma_{5}M\sin2\phi}{p^{2}-M^{2}+i\epsilon}.
\end{eqnarray}
If one takes the limit $p^{0}\rightarrow \infty$ in this relation and use the same type of reasoning as above, we find
\begin{eqnarray}
\int d^{4}x e^{ip(x-y)}i\delta(x^{0}-y^{0})\{i\partial_{x^{0}}n^{c}(t,\vec{x}),\bar{n}(t,\vec{y})\}=(-i)\gamma_{5}M\sin2\phi
\end{eqnarray}
or
\begin{eqnarray}
i\delta(x^{0}-y^{0})\{i\partial_{x^{0}}n^{c}(t,\vec{x}),\bar{n}(t,\vec{y})\}=(-i)\gamma_{5}M\sin2\phi\delta^{4}(x-y),
\end{eqnarray}
as well as
\begin{eqnarray}
&&\int d^{4}x e^{ip(x-y)}
\langle T (i\partial_{x^{0}})^{2}n^{c}(x)\bar{n}(y)\rangle= \frac{(\vec{p}^{2}+M^{2})(-i)\gamma_{5}M\sin2\phi}{p^{2}-M^{2}+i\epsilon}.
\end{eqnarray}
This last term is equivalently written as 
\begin{eqnarray}
&&\int d^{4}x e^{ip(x-y)}
\langle T \left([-(\Box+M^{2})+(-\partial_{k}\partial_{k}+M^{2})]n^{c}(x)\right)\bar{n}(y)\rangle\nonumber\\
&=&\int d^{4}x e^{ip(x-y)}
\langle T [(-\partial_{k}\partial_{k}+M^{2})n^{c}(x)]\bar{n}(y)\rangle\nonumber\\
&=&\left(\vec{p}^{2}+M^{2}\right)\int d^{4}x e^{ip(x-y)}
\langle T n^{c}(x)\bar{n}(y)\rangle\nonumber\\
&=& \frac{(\vec{p}^{2}+M^{2})(-i)\gamma_{5}M\sin2\phi}{p^{2}-M^{2}+i\epsilon}.
\end{eqnarray}
and thus 
\begin{eqnarray}
\int d^{4}x e^{ip(x-y)}
\langle T n^{c}(x)\bar{n}(y)\rangle
&=& \frac{(-i)\gamma_{5}M\sin2\phi}{p^{2}-M^{2}+i\epsilon},
\end{eqnarray}
namely, we come back to the starting relation \eqref{T-product}. In this 
derivation, we used the equation of motion $(\Box+M^{2})n^{c}(x)=0$ 
suggested by $n^{c}(x)=\cos\phi N^{c}(x)-\gamma_{5}\sin\phi N(x)$.

We thus obtain
\begin{eqnarray}\label{comm-1}
&&\delta(x^{0}-y^{0})\{n^{c}(t,\vec{x}),\bar{n}(t,\vec{y})\}=0,\\
&&\delta(x^{0}-y^{0})\{i\partial_{x^{0}}n^{c}(t,\vec{x}),\bar{n}(t,\vec{y})\}=\delta^{4}(x-y)
(-1)\gamma_{5}M\sin2\phi,\nonumber
\end{eqnarray}
where the second relation is a novel anti-commutation relation from a naive canonical point of view,
recalling that
$\partial_{x^{0}}n^{c}(t,\vec{x})=C\dot{\overline{n}}^{T}(t,\vec{x})$,
such that
\begin{eqnarray}\label{novel-commutator1}
i\{\dot{\overline{n}}(t,\vec{x}), \bar{n}(t,\vec{y})\}= - C^{-1}\gamma_{5}\epsilon_{5}\delta(\vec{x}-\vec{y}) 
\end{eqnarray}
by noting $M\sin2\phi=\epsilon_{5}$.
\\

Similarly, starting with the correlation function
\begin{eqnarray}\label{T-product2}
\int d^{4}x e^{ip(x-y)}\langle T n(x)\overline{n^{c}}(y)\rangle
= \frac{i\gamma_{5}M\sin2\phi}{p^{2}-M^{2}+i\epsilon},
\end{eqnarray}
and repeating similar analyses, we obtain the equal-time anti-commutators  
 \begin{eqnarray}\label{comm-2}
&&\delta(x^{0}-y^{0})\{n(t,\vec{x}),\overline{n^{c}}(t,\vec{y})\}=0,\\
&&\delta(x^{0}-y^{0})\{i\partial_{x^{0}}n(t,\vec{x}),\overline{n^{c}}(t,\vec{y})\}=\delta^{4}(x-y)
\gamma_{5}M\sin2\phi.\nonumber
\end{eqnarray}
The last relation implies the novel anti-commutation relation
\begin{eqnarray}\label{novel-commutator2}
i\{\dot{n}(t,\vec{x}),n(t,\vec{y})\}=-\gamma_{5}C\epsilon_{5}\delta(\vec{x}-\vec{y}).
\end{eqnarray}

We also have a correlation function from \eqref{12}
\begin{eqnarray}
\int d^{4}x e^{ip(x-y)}\langle T^{\star} n(x)\overline{n}(y)\rangle
= \frac{i(\pslash+ M\cos2\phi)}{p^{2}-M^{2}+i\epsilon},
\end{eqnarray}
which, using BJL prescription, leads to
\begin{eqnarray}\label{T-product3}
\int d^{4}x e^{ip(x-y)}\langle T n(x)\overline{n}(y)\rangle
= \frac{i(\pslash+ M\cos2\phi)}{p^{2}-M^{2}+i\epsilon}.
\end{eqnarray}
By multiplying both sides by $-ip^{0}$, we have
\begin{eqnarray}
&&\int d^{4}x e^{ip(x-y)}\partial_{x^{0}}\langle T n(x)\overline{n}(y)\rangle\nonumber\\
&&=\int d^{4}x e^{ip(x-y)}[\delta(x^{0}-y^{0})\{n(t,\vec{x}),\overline{n}(t,\vec{y})  \}+\langle T \partial_{x^{0}}n(x)\overline{n}(y)\rangle]\nonumber\\
&&=\frac{p^{0}(\pslash+ M\cos2\phi)}{p^{2}-M^{2}+i\epsilon}.
\end{eqnarray}
Taking the limit $p^{0}\rightarrow\infty$, we find 
\begin{eqnarray}
\delta(x^{0}-y^{0})\{n(t,\vec{x}),\overline{n}(t,\vec{y})\}=\gamma^{0}\delta^{4}(x-y)
\end{eqnarray}
and 
\begin{eqnarray}
\int d^{4}x e^{ip(x-y)}\langle T \partial_{x^{0}}n(x)\overline{n}(y)\rangle
=\frac{p^{0}(p_{k}\gamma^{k}+ M\cos2\phi)+\gamma^{0}(\vec{p}^{2}+M^{2})}{p^{2}-M^{2}+i\epsilon}.
\end{eqnarray}
Multiplying both sides of this last relation by $-p^{0}$, we have
\begin{eqnarray}
&&\int d^{4}x e^{ip(x-y)}[\delta(x^{0}-y^{0})\{\partial_{x^{0}}n(t,\vec{x}),\overline{n}(t,\vec{y})\}+\langle T \partial^{2}_{x^{0}}n(x)\overline{n}(y)\rangle\}\nonumber\\
&&=\frac{-i(p^{0})^{2}(p_{k}\gamma^{k}+ M\cos2\phi)-ip^{0}\gamma^{0}(\vec{p}^{2}+M^{2})}{p^{2}-M^{2}+i\epsilon}.
\end{eqnarray}
The consideration with $p^{0}\rightarrow \infty$ gives 
\begin{eqnarray}
[\delta(x^{0}-y^{0})\{\partial_{x^{0}}n(t,\vec{x}),\overline{n}(t,\vec{y})\}=(-\partial_{k}\gamma^{k}-iM\cos2\phi)\delta(x-y)
\end{eqnarray}
and 
\begin{eqnarray}\label{48}
&&\int d^{4}x e^{ip(x-y)}\langle T \partial^{2}_{x^{0}}n(x)\overline{n}(y)\rangle\nonumber\\
&&=\frac{-i(p^{0})^{2}(p_{k}\gamma^{k}+ M\cos2\phi)-ip^{0}\gamma^{0}(\vec{p}^{2}+M^{2})}{p^{2}-M^{2}+i\epsilon}+i(p_{k}\gamma^{k}+ M\cos2\phi)\nonumber\\
&&=-(\vec{p}^{2}+M^{2})\frac{i\pslash+M\cos2\phi}{p^{2}-M^{2}+i\epsilon}.
\end{eqnarray}
Using the equation of motion for $n(x)$, the last relation \eqref{48} leads to 
\begin{eqnarray}
-(\vec{p}^{2}+M^{2})\int d^{4}x e^{ip(x-y)}\langle T n(x)\overline{n}(y)\rangle
=-(\vec{p}^{2}+M^{2})\frac{i\pslash+M\cos2\phi}{p^{2}-M^{2}+i\epsilon}
\end{eqnarray}
which in turn gives the starting expression \eqref{T-product3}.
We thus summarize the derived anti-commutators as
\begin{eqnarray}\label{comm-3}
\delta(x^{0}-y^{0})\{n(t,\vec{x}),\overline{n}(t,\vec{y})\}&=&\gamma^{0}\delta^{4}(x-y),\nonumber\\
\delta(x^{0}-y^{0})\{\partial_{x^{0}}n(t,\vec{x}),\overline{n}(t,\vec{y})\}&=&(-\gamma^{k}\partial_{k} -iM\cos2\phi)\delta^{4}(x-y).
\end{eqnarray}

The novel canonical anti-commutation relations, namely, the second relations in \eqref{comm-1} and \eqref{comm-2}, arise from the unconventional correlation functions \eqref{T-product} and \eqref{T-product2}, respectively, which describe the baryon number violating effects in the absence of neutron-antineutron oscillation. The deviation of $M\cos2\phi$ from $M$ for $\phi\neq 0$ in the second relation of \eqref{comm-3} is also a novel
anti-commutation relation.  

In passing, we mention why no novel anti-commutation relations appear in the baryon-number conserving theory such as \eqref{NJL}.  If one starts with \eqref{nambu},
\begin{eqnarray}
\int d^{4}x e^{ip(x-y)}\langle T n(x)n(y)\rangle=0,
\end{eqnarray}
one obtains the relations by multiplying $p^{0}$ and taking the limit $p^{0}\rightarrow \infty$,
\begin{eqnarray}
&&\{n(t, \vec{x}), n(t,\vec{y})\}=0, \nonumber\\
&&\{\dot{n}(t, \vec{x}), n(t,\vec{y})\}=0, \nonumber\\
&&\{\ddot{n}(t, \vec{x}), n(t,\vec{y})\}=0,  \nonumber\\
&& ..... . 
\end{eqnarray}  
Namely, no novel anti-commutators arise.

\subsection{Canonical operator analysis}
We work with the explicit effective Lagrangian \eqref{1} with $\epsilon_{1}=0$,
\begin{eqnarray}
{\cal L}&=&\overline{n}(x)i\gamma^{\mu}\partial_{\mu}n(x) - m\overline{n}(x)n(x)\nonumber\\
&-&\frac{1}{2}\epsilon_{5}[n^{T}(x)C\gamma_{5}n(x) - \overline{n}(x)C\gamma_{5}\overline{n}^{T}(x)],
\end{eqnarray}
which is violating C and P ("$i\gamma^{0}$-parity") separately  but preserves CP. This Lagrangian is  also invariant under the "$\gamma^{0}$-parity", whose implication has been already discussed.
This Lagrangian is re-written as
\begin{eqnarray}\label{hamiltonian}
{\cal L}&=&\overline{n}(x)i\gamma^{\mu}\partial_{\mu}n(x) - m\overline{n}(x)n(x)\nonumber\\
&&-\frac{1}{2}\epsilon_{5}[n^{T}(x)C\gamma_{5}n(x) - \overline{n}(x)C\gamma_{5}\overline{n}^{T}(x)]\nonumber\\
&=&\overline{n}(x)i\gamma^{0}\partial_{0}n(x)-\{-
\overline{n}(x)i\gamma^{k}\partial_{k}n(x) + m\overline{n}(x)n(x)\nonumber\\
&&+\frac{1}{2}\epsilon_{5}[n^{T}(x)C\gamma_{5}n(x) - \overline{n}(x)C\gamma_{5}\overline{n}^{T}(x)]\}\nonumber\\
&\equiv&\Pi_{n}(x)\partial_{0}n(x)-{\cal H}
\end{eqnarray}
where
\begin{eqnarray}
\Pi_{n}(x)&=&\overline{n}(x)i\gamma^{0}\nonumber\\
&=&i n^{\dagger}(x),\nonumber\\
{\cal H}(\Pi_{n}, n)&=&-
\overline{n}(x)i\gamma^{k}\partial_{k}n(x) + m\overline{n}(x)n(x)\nonumber\\
&&+\frac{1}{2}\epsilon_{5}[n^{T}(x)C\gamma_{5}n(x) - \overline{n}(x)C\gamma_{5}\overline{n}^{T}(x)]  
\end{eqnarray}
and the canonical anti-commutators are
\begin{eqnarray}  
\{n(t,\vec{x}), \Pi_{n}(t, \vec{y})\} &=&\{n(t,\vec{x}), in^{\dagger}(t, \vec{y})\}=i\delta(\vec{x}-\vec{y}),\nonumber\\
\{n(t,\vec{x}), n(t, \vec{y})\} &=&0,\nonumber\\
\{n^{\dagger}(t,\vec{x}), n^{\dagger}(t, \vec{y})\} &=&0.
\end{eqnarray}
A salient feature of the present scheme is that 
\begin{eqnarray} 
i\partial_{t}n(t, \vec{x})&=&[n(t, \vec{x}), \int d^{3}y {\cal H}]\nonumber\\
&=&-\gamma^{0}i\gamma^{k}\partial_{k}n(t, \vec{x}) + m\gamma^{0}n(t, \vec{x}) - \epsilon_{5} \gamma^{0}C\gamma_{5}\overline{n}^{T}(t, \vec{x}),  
\end{eqnarray}
which implies that 
\begin{eqnarray} 
i\{\partial_{t}n(t, \vec{x}), n(t, \vec{y})\}
&=&\{- \epsilon_{5} \gamma^{0}C\gamma_{5}\overline{n}^{T}(t, \vec{x}),  n(t, \vec{y})\}\nonumber\\
&=&- \epsilon_{5} \gamma^{0}C\gamma_{5}\gamma^{0}\delta(\vec{x}-\vec{y})
\nonumber\\
&=&- \epsilon_{5}C\gamma_{5} \delta(\vec{x}-\vec{y}) ,
\end{eqnarray}
in agreement with \eqref{novel-commutator2}.  Other novel commutators such as \eqref{novel-commutator1} are similarly established.

A drawback of the present formulation, which is very simple, is that a direct connection with the transition amplitude \eqref{T-product2} is not very transparent.
A more detailed operator formulation in terms of creation and annihilation operators including an analysis which clarifies the  connection of our Bogoliubov transformation with the transformation used by Nambu and Jona-Lasinio~\cite{nambu} will be given elsewhere.

\section{Discussion and conclusion}
 
We have shown the appearance of novel canonical anti-commutation relations in the presence of baryon number violating terms, which, to our knowledge, has not been discussed before. This 
modification was recognized by the BJL method first. But after a careful analysis of equations of motion, we have shown that this modification is also understood in the operator formalism. Interestingly, this modification has an important physical implication, namely, the neutron decays through the baryon number violating channels into two modes, even in the case where  the oscillation between the neutron and antineutron is absent due to the degenerate  Majorana fermion masses.

The novel nonzero propagator \eqref{15}, which is the starting point of our analysis, is a direct consequence of the change of vacuum which we mentioned in Section 2. Indeed, the propagator is considered as expectation value on the true ground state of the theory, which is the vacuum of the quasiparticles $N$, as one can see from \eqref{propagator}. As such, it includes the effect of the condensation of neutron pairs, which reflects the baryon number violation. 

Finally, an analysis similar to the present one is in principle applicable to the seesaw mechanism for neutrino masses, although the specific choice of the parameter $\epsilon_{1}=0$, for example, is not relevant there~\cite{FT, FT2}. 
\\

The support of the Academy of Finland under the Projects no. 136539 and 272919 is gratefully acknowledged.

\end{document}